\documentstyle[11pt,AATS,epsf]{article}	% DO NOT TOUCH

\begin{document}	% DO NOT TOUCH 

%-----------------------------------------------------------------------
%       Paper Title 
%-----------------------------------------------------------------------
% Enter the title of the paper. Please use mixed cases.
\title{Horizontal-Branch Morphology as an Age Indicator}

%-----------------------------------------------------------------------
%	Author(s) names and affiliation
%-----------------------------------------------------------------------
% Enter the authors followed by their affiliations.  The \author and
% \affil commands may appear multiple times as necessary. List each
% author by giving the first name or initials first followed by the last
% name.  Authors with the same affiliations should be grouped together,
% with the names separated by commas (no &, no 'and').
\author{Young-Wook Lee, Suk-Jin Yoon and Soo-Chang Rey}
\affil{Center for Space Astrophysics \& Department of Astronomy, Yonsei University, 
Shinchon 134, Seoul 120-749, Korea}
\author{Brian Chaboyer}
\affil{Department of Physics and Astronomy, 6127 Wilser Lab., Dartmouth College, Hanover,
NH 03755-3528, USA}

%-----------------------------------------------------------------------
%	Contact information
%-----------------------------------------------------------------------
% Please include a comment (a line which starts with a %) with the name
% and email address of a contact person in case the editors experience
% difficulties with your manuscript.
% Soo-Chang Rey, screy@csa.yonsei.ac.kr

%-----------------------------------------------------------------------
%                              Abstract
%-----------------------------------------------------------------------
% Type abstract in the space below, between the "\begin{abstract}" and
% "\end{abstract}" lines. No blank line after the "\begin{abstract}"
% line! 
\begin{abstract}
%Surface temperature distribution of horizontal-branch (HB) stars is very 
%sensitive to age in old stellar systems, which makes it an attractive age
%indicator. 
We present our recent revision of model constructions for the horizontal-branch (HB)
morphology of globular clusters, which suggests the HB morphology is more 
sensitive to age compared to our earlier models. We also present our 
high precision CCD photometry for the classic second parameter pair
M3 and M13. The relative age dating based on this photometry indicates that 
M13 is indeed older than M3 by 1.7 Gyr. This is consistent with the age 
difference predicted from our new models, which provides a further support 
that the HB morphology is a reliable age indicator in most population II
stellar systems.
\end{abstract}

%-----------------------------------------------------------------------
%                             Main Body
%-----------------------------------------------------------------------
% After the abstract comes the main body of the manuscript. Use \section
% to label various sections. You may also use \subsection. Please
% use mixed cases in the sections' titles. Sections and subsections will
% be numbered automatically.  

\section{Introduction}
Because the horizontal-branch (HB) stars in globular clusters are much brighter 
than main-sequence (MS) stars, the interpretation of HB morphology in terms of 
relative age differences would be of great value in the study of distant stellar 
populations where the MS turnoff is fainter than the detection limit. In this 
paper, we report our progress in the use of HB as a reliable age indicator.

\section{New HB Population Models}
Some seven years ago, Lee et al. (1994, hereafter LDZ) have concluded that age 
is the most natural candidate for the global second parameter, because other 
candidates can be ruled out from the observational evidence, while supporting 
evidence do exist for the age hypothesis. Although this conclusion is generally 
accepted in the community, there is still some debate about this issue (see Table 
1 of Lee et al. 1999). Among others, critics are arguing that the relative age 
differences inferred from the MS turnoffs are often too small compared to the 
amounts predicted from the HB models for the several second parameter pairs, 
including the well-known M3 and M13 pair. 

We found, however, several recent developments can affect the relative age 
dating technique from HB morphology. First of all, there is now a reason to 
believe that absolute age of the oldest Galactic globular clusters is reduced to 
about 12 Gyr, as suggested by new Hipparcos distance calibration and other 
improvements in stellar models (Reid 1997; Gratton et al. 1997; Chaboyer et al. 1998). 
As LDZ already demonstrated in their paper, 
this has a strong impact in the relative age estimation from HB morphology. 
Also, it is now well established that $\alpha$ elements are enhanced in halo 
populations ([$\alpha$/Fe] = 0.4). Finally, Reimers'(1975) empirical mass-loss law suggests 
more mass-loss at larger ages. The result of this effect was also presented in 
LDZ, but unfortunately most widely used diagram (their Fig. 7) is the one 
based on fixed mass-loss. We found all of the above effects make the HB 
morphology to be more sensitive to age (see Lee \& Yoon 2001 for details). 
Therefore, as illustrated in Figure 1 , 
now the required age difference is much reduced compared to Figure 7 of LDZ. 
Now, only 1.1 Gyr of age difference, rather than 2 Gyr, is enough to explain 
the systematic shift of the HB morphology between the inner and outer halo 
clusters. Also, to within the observational uncertainty, age difference of about 
1.5-2 Gyr is now enough to explain the observed difference in HB morphology 
between the remote halo clusters (Pal 3, Pal 4, Pal 14, and Eridanus) and M3. 
These values are consistent with the recent relative age datings both from the 
HST and high-quality ground-based data.

\begin{figure}[hb]	% the figure environment
\plotfiddle{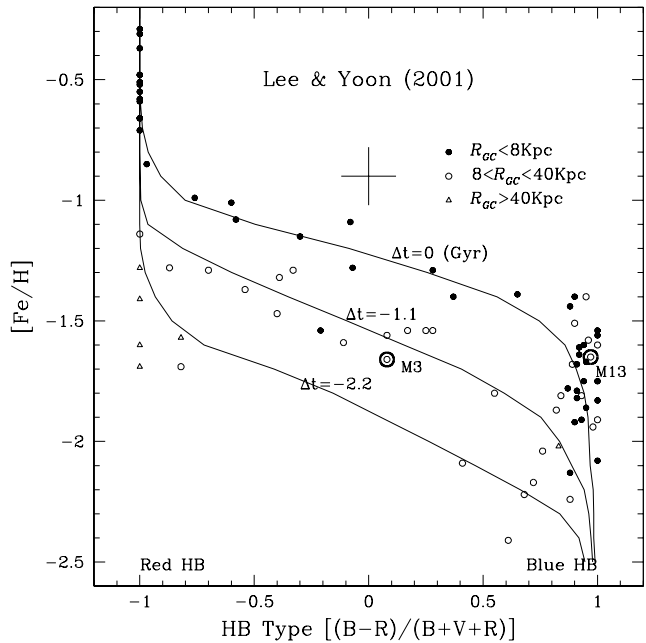}{8cm}{0}{43}{43}{-130}{-67}
\caption{New HB population models are more sensitive to age compared to our 
earlier models. $\Delta$t = 0 corresponds to the mean age of the inner halo (R $<$ 8 
Kpc) clusters, and the relative ages are in Gyr.}
\end{figure}

\section{Case of M3 and M13}
We obtained high quality color-magnitude data for the classic second parameter globular 
clusters M3 and M13. The clusters were observed during the same nights with 
the same instruments (MDM 2.4m), allowing us to determine the accurate 
relative ages. From the color difference method between the turnoff and the base 
of the red giant-branch (RGB), we now confirm M13 is indeed 1.7$\pm$0.5 Gyr 
older than M3 (Fig. 2; see Rey et al. 2001 for details). This is consistent with 
the age difference predicted from our new HB models (see Fig. 1), which 
provides a further support that the HB morphology is a reliable age indicator in 
most population II stellar systems.

\begin{figure}[hb]	% the figure environment
\plotfiddle{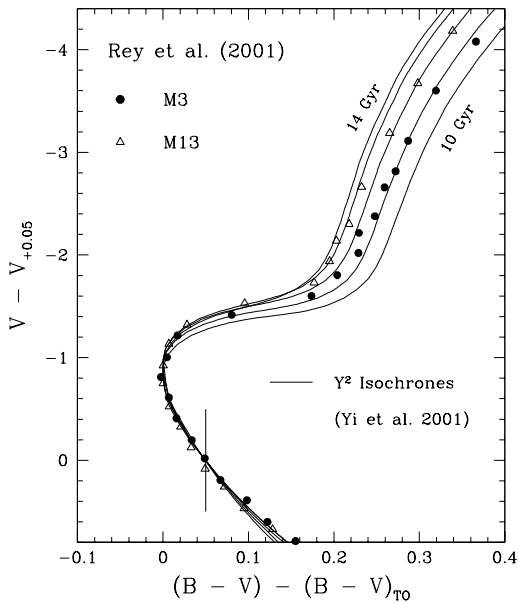}{7.3cm}{0}{46}{46}{-150}{-98}
\caption{ Fiducial sequence for M3 is compared with that for M13 following the 
prescription of VandenBerg et al. (1990). Note a separation between the two 
clusters' RGBs, indicating an age difference.}
\end{figure}

%-----------------------------------------------------------------------
%                             References
%-----------------------------------------------------------------------
% List your references below between the \begin{references} and
% \end{references} commands. Each reference should begin with a
% \reference command. Observe the following order when listing
% bibliographical information for each reference:  author name(s),
% publication year, journal name, volume, and page number for
% articles.  See the User's Guide for a list of macros to represent
% journals. 

\acknowledgments

\end{document}